\documentclass{elsart}
\usepackage{graphicx}
\usepackage{latexsym}
\begin{document}
\begin{frontmatter}
\hyphenation{equa-tion how-ever dom-in-ant}
\title{Hybrid mesons from lattice QCD}
\author{Colin Morningstar}
\address{Florida International University, Miami, Florida, USA}
\date{7 August 2000}
\begin{abstract}
  Recent lattice simulation studies of heavy-quark hybrid mesons in which
the quark and antiquark are bound together by an excited gluon field are
summarized.
\end{abstract}
\end{frontmatter}
Hybrid mesons are states in which a valence quark and an antiquark are
bound by an excited gluon field.  Interest in such states has been
recently sparked by observations of resonances with exotic $1^{-+}$
quantum numbers\cite{E852}.  In fact, the proposed Hall D at Jefferson
Lab will be devoted to the search for hybrid mesons.  Although our
understanding of these states remains meager, recent lattice simulations
have shed some light on their nature.  In this talk, I focus on studies
of hybrid mesons containing heavy quarks, although I conclude with brief
comments on light-quark hybrid-meson simulations.

One expects that a heavy-quark meson can be treated similar to a diatomic
molecule: the slow valence heavy quarks correspond to the nuclei and the fast
gluon and light sea quark fields correspond to the electrons\cite{hasenfratz}.
First, the quark $Q$ and antiquark $\overline{Q}$ are treated as
static color sources and the energy levels of the fast degrees of freedom
are determined as a function of the $Q\overline{Q}$ separation $r$,  each such
energy level defining an adiabatic surface or potential.  The motion of the
slow heavy quarks is then described in the leading Born-Oppenheimer (LBO)
approximation by the Schr\"odinger equation using each of these potentials.
Conventional quarkonia are based on the lowest-lying potential; hybrid
quarkonium states emerge from the excited potentials.  

The validity of such a simple physical picture relies on the smallness
of higher-order spin, relativistic, and retardation effects and mixings
between states based on different adiabatic surfaces.  Recently, it was
demonstrated that retardation and mixing effects do not spoil the leading
Born-Oppenheimer approximation for the (quenched) low-lying conventional
and hybrid spectrum\cite{jkm}.  In Ref.~\cite{drummond}, the introduction
of the heavy-quark spin was shown to lead to significant level shifts but
{\em not} to significant mixings between states based on different
potentials.  In this talk, the leading Born-Oppenheimer approximation
is described, the findings of Refs.~\cite{jkm} and \cite{drummond} are
summarized, and the expected (but currently unknown) impact of sea quark
effects on the LBO spectrum calculation is discussed.

\begin{figure}
\begin{center}
\resizebox{0.85\textwidth}{!}{\includegraphics[bb= 167 354 488 582]{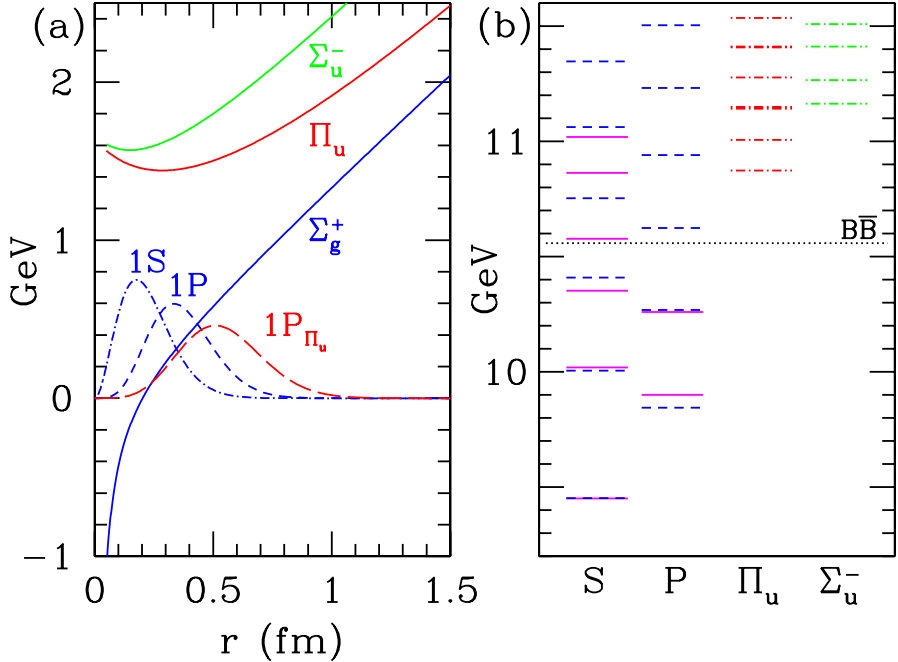}}
\caption[figwf]{(a) Static potentials and radial probability densities
 against quark-antiquark separation $r$ for $r_0^{-1}=450$ MeV. 
 (b) Spin-averaged $\bar{b}b$ spectrum
 in the LBO approximation (light quarks neglected).  Solid lines
 indicate experimental measurements.  Short dashed lines indicate 
 the $S$ and $P$ state masses obtained using the $\Sigma_g^+$ potential
 with $M_b=4.58$ GeV. Dashed-dotted
 lines indicate the hybrid quarkonium states obtained from the $\Pi_u$
 $(L=1,2,3)$ and $\Sigma_u^-$ $(L=0,1,2)$ potentials.
\label{fig:LBO}}
\end{center}
\end{figure}

The spectrum of the fast gluon field in the presence of a static
quark-antiquark pair has been determined in lattice studies\cite{earlier}.
The three lowest-lying levels are shown in Fig.~\ref{fig:LBO}.  Due
to computational limitations, sea quark effects have been neglected in
these calculations; their expected impact on the hybrid meson spectrum
will be discussed later.  The levels in Fig.~\ref{fig:LBO} are labeled
by the magnitude $\Lambda$ of the projection of the total angular momentum
${\bf J}_g$ of the gluon field onto the molecular axis, and by $\eta=\pm 1$,
the symmetry under the charge conjugation combined with spatial inversion
about the midpoint between the $Q$ and $\overline{Q}$.  States with
$\Lambda=0,1,2,\dots$ are denoted by $\Sigma, \Pi, \Delta, \dots$,
respectively.  States which are even (odd) under the above-mentioned
$CP$ operation are denoted by the subscripts $g$ ($u$).  An additional
$\pm$ superscript for the $\Sigma$ states refers to even or odd symmetry
under a reflection in a plane containing the molecular axis.  The
potentials are calculated in terms of the hadronic scale parameter
$r_0$; in Fig.~\ref{fig:LBO}, $r_0^{-1}=450$ MeV has been assumed. 

The LBO spectrum is obtained by solving the radial Schr\"odinger
equation.  Results for the LBO spectrum of conventional $\overline{b}b$
and hybrid $\overline{b}gb$ states are shown in Fig.~\ref{fig:LBO}. 
Below the $\overline{B}B$ threshold, the LBO results are in very good
agreement with the spin-averaged experimental measurements of bottomonium
states.   Above the threshold, agreement with experiment is lost, suggesting
significant corrections either from mixing and other higher-order effects or
(more likely) from light sea quark effects.  Note from the radial probability
densities shown in Fig.~\ref{fig:LBO} that the size of the hybrid state is
large in comparison with the conventional $1S$ and $1P$ states. 

\begin{figure}
\begin{center}
\resizebox{0.65\textwidth}{!}{
  \includegraphics[bb= 28 144 572 552]{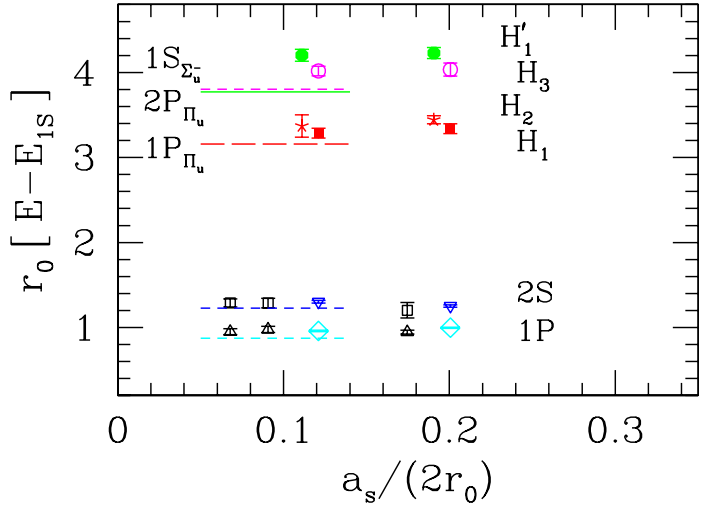}}
\caption[figscaling]{
  Simulation results from Ref.~\protect\cite{jkm} for the level splittings
  (in terms of $r_0$ and with respect to the $1S$ state) against the lattice
  spacing $a_s$. Results from Ref.~\protect\cite{Wilson} using an NRQCD
  action with higher-order corrections are shown as $\Box$ and
  $\bigtriangleup$.  The horizontal lines show the LBO predictions.
\label{fig:scaling}}
\end{center}
\end{figure}

The importance of retardation and leading-order mixings between states
based on different adiabatic potentials can be tested by comparing the LBO
level splittings with those determined from meson simulations using a
leading-order non-relativistic (NRQCD) heavy-quark action.  Such a test was
carried out in Ref.~\cite{jkm}.  The NRQCD action included only a covariant
temporal derivative and the leading kinetic energy operator (with two other
operators to remove lattice spacing errors).  The only difference between the
leading Born-Oppenheimer Hamiltonian and the lowest-order NRQCD Hamiltonian
was the $\mbox{\boldmath{$p\!\cdot\!A$}}$ coupling between the quark color
charge in motion and the gluon field.   The level splittings (in terms of
$r_0$ and with respect to the $1S$ state) of the conventional $2S$ and $1P$
states and four hybrid states were compared (see Fig.~\ref{fig:scaling})
and found to agree within $10\%$, strongly supporting the validity of the
leading Born-Oppenheimer picture.  The meson operators used are listed
in Table~\ref{table:mesonops} and are described in much more detail
in Ref.~\cite{jkm}.

\begin{table}[t]
\begin{center}
\begin{tabular}{|c|l|c|c|}\hline
 $J^{PC}$ & & Degeneracies & Operator \\ \hline \hline
 $0^{-+}$ & $S$ wave & $1^{--}$ &$\tilde\chi^\dag\ \left[\tilde\Delta^{(2)}
     \right]^p\ \tilde\psi$ \\
 $1^{+-}$ & $P$ wave & $0^{++},1^{++},2^{++}$ &$\tilde\chi^\dag
    \ \tilde{\mbox{\boldmath{$\Delta$}}} \ \tilde\psi$ \\
 $1^{--}$ & $H_1$ hybrid & $0^{-+},1^{-+},2^{-+}$ &
    $\tilde\chi^\dag\ \tilde{\bf B}
       \left[\tilde\Delta^{(2)}\right]^p\ \tilde\psi$ \\
 $1^{++}$ & $H_2$ hybrid & $0^{+-},1^{+-},2^{+-}$ & $\tilde\chi^\dag\ 
      \tilde{\bf B}\!\times\!\tilde{\mbox{\boldmath{$\Delta$}}}
      \ \tilde\psi$ \\
 $0^{++}$ & $H_3$ hybrid & $1^{+-}$ & $\tilde\chi^\dag\ \tilde{\bf B}
     \!\cdot\!\tilde{\mbox{\boldmath{$\Delta$}}}\ \tilde\psi$ \\ \hline
\end{tabular}
\end{center}
\caption[tabtwo]{
  The meson operators used in Ref.~\protect\cite{jkm}.  Note that
  in the $0^{-+}$ and $1^{--}$ sectors, four operators were used
  by taking $p=0,1,2,3$.  The corresponding $S=1$ states, degenerate
  for the spin-independent NRQCD action used, are also listed.
  \label{table:mesonops}}
\end{table}

\begin{figure}
\begin{center}
\resizebox{0.65\textwidth}{!}{
  \includegraphics[bb= 16 50 507 410]{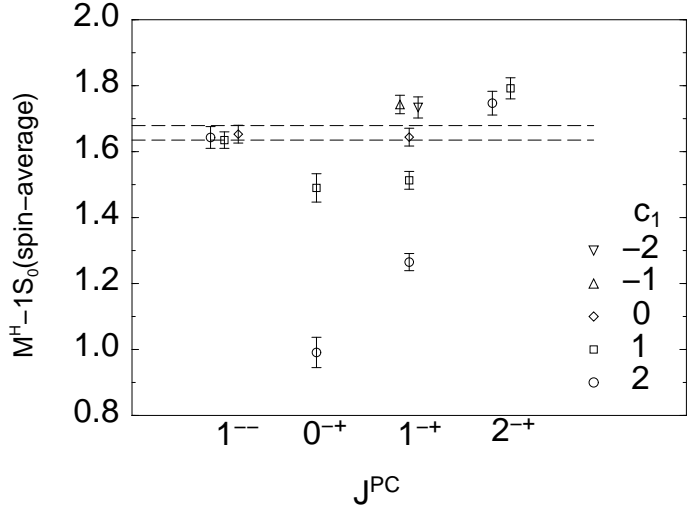}}
\caption[figspin]{
  Variation of the hybrid masses (in GeV) with $c_1$ (from 
  Ref.~\protect\cite{drummond}).
\label{fig:spinshifts}}
\end{center}
\end{figure}

\begin{figure}[b]
\begin{center}
\resizebox{0.85\textwidth}{!}{
  \includegraphics[bb= 0 0 377 87]{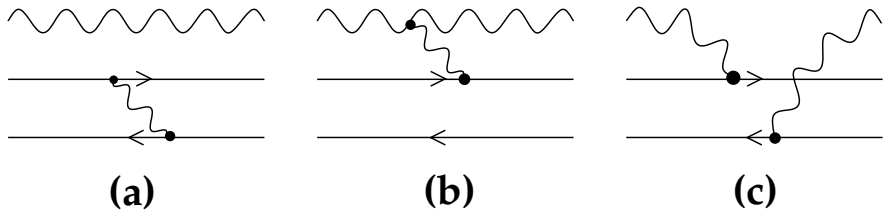}}
\caption[feyn]{
  Bag model diagrams studied in Ref.~\protect\cite{drummond}.  A wavy
  line indicates a gluon, and the two straight lines with arrows show the
  quark and antiquark.
\label{fig:feyndiags}}
\end{center}
\end{figure}

The question of whether or not quark spin interactions spoil the 
validity of the Born-Oppenheimer picture for heavy-quark hybrids
has been addressed in Ref.~\cite{drummond}.  Simulations of several
hybrid mesons using an NRQCD action including the spin interaction
$-c_1 \mbox{\boldmath{$\sigma\!\cdot\! B$}}/2M_b$ and neglecting
light sea quark effects were carried out; the resulting splittings
are shown in Fig.~\ref{fig:spinshifts} for various values of the
coupling $c_1$ (which is expected to be near unity) and are surprisingly
large.  However, the authors of Ref.~\cite{drummond} suggest that these
splittings do {\em not} signal a breakdown of the Born-Oppenheimer picture.
First, they claim that no significant mixing of their non-exotic $0^{-+}$,
$1^{--}$, and $2^{-+}$ hybrid meson operators with conventional states was
observed; unfortunately, this claim is not convincing since a correlation
matrix analysis was not used and the effective masses associated with
these correlators were not shown.  Secondly, the authors argue that
calculations using the bag model support their suggestion.  Contributions
to the spin splittings in Fig.~\ref{fig:spinshifts} from the three diagrams
shown in Fig.~\ref{fig:feyndiags} were estimated in the bag model.
Graph (a) is a weak effect (producing level shifts of order 1 MeV or less)
since it involves the $Q\overline{Q}$ wavefunction near the origin which is
suppressed in hybrid states.   Graph (b) is the dominant contribution
(producing level shifts consistent with those shown in 
Fig.~\ref{fig:spinshifts}) since it depends on the quark wavefunction in the
bulk; significant contributions from graph (b) do not necessarily spoil
the Born-Oppenheimer picture.  Graph (c) represents mixings between states
based on different adiabatic surfaces and large contributions from such
diagrams would signal a breakdown of the Born-Oppenheimer picture; however,
bag model estimates of this mixing amplitude are very small of 
order $10^{-4}$.  These facts are not conclusive evidence that heavy-quark
spin effects do not spoil the Born-Oppenheimer picture, but they are highly
suggestive.

The dense spectrum of hybrid states shown in Fig.~\ref{fig:LBO} neglects
the effects of light sea quark-antiquark pairs.  In order to include
these effects in the LBO, the adiabatic potentials must be determined
fully incorporating the light quark loops.  Such computations using
lattice simulations are very challenging, and to date, high precision
results for these potentials for separations out to 2~fm are not yet
available.  For separations below 1 fm, the $\Sigma_g^+$ and $\Pi_u$
potentials change very little\cite{bali} from the behavior shown in 
Fig.~\ref{fig:LBO}, suggesting that a few of the lowest-lying hybrid
states may exist as well-defined resonances. However, for $Q\overline{Q}$
separations greater than 1~fm, the adiabatic surfaces should change
dramatically from the behavior shown in Fig.~\ref{fig:LBO}; instead of
increasing indefinitely, they should eventually level off since the
static $\overline{Q}gQ$ state can undergo fission into two separate
$\overline{Q}q$ color singlets, where $q$ is a light quark.  The
precise behavior of these potentials due to configuration mixing
with $\overline{B}B$ mesons in the transition region is unknown; in fact,
it is not even known if a fission barrier occurs.  Clearly, such
potentials cannot support the plethora of conventional and hybrid states
shown in Fig.~\ref{fig:LBO}; the formation of bound states and resonances
substantially extending over 1~fm seems unlikely.  Whether or not the
light sea quark-antiquark pairs spoil the Born-Oppenheimer picture is
currently unknown.  Future unquenched simulations should help to answer
this question.  I remain hopeful that the simple physical picture provided
by the Born-Oppenheimer expansion for both the low-lying conventional and
hybrid heavy-quark mesons will survive the introduction of the light sea
quark effects.  I should mention once again that the discrepancies of the
spin-averaged LBO predictions with experiment above the $B\overline{B}$
threshold seen in Fig.~\ref{fig:LBO} are possibly caused by our neglect
of light sea quark-antiquark pairs. 

A summary of recent light-quark and charmonium $1^{-+}$ hybrid mass
calculations is presented in Table~\ref{table:light}.  With the exception
of Ref.~\cite{lasch98}, all results neglect light sea quark loops.  The
introduction of two flavors of dynamical quarks in Ref.~\cite{lasch98}
yielded little change to the hybrid mass, but this finding should not be
considered definitive due to uncontrolled systematics (unphysically large
quark masses, inadequate treatment of resonance properties in finite volume,
{\it etc.}).  All estimates of the light quark hybrid mass are near 
2.0 GeV, well above the experimental candidates found in the range
1.4-1.6 GeV.  Perhaps sea quark effects will resolve this discrepancy,
or perhaps the observed states are {\em not} hybrids.  Some authors have
suggested that they may be four quark $\bar{q}\bar{q}qq$ states.  Clearly,
there is still much to be learned about these exotic QCD resonances.

\begin{table}[t]
\begin{center}
\begin{tabular}{|ll|c|l||ll|l|}\hline
 \multicolumn{4}{|c||}{Light quark $1^{-+}$}
 &\multicolumn{3}{|c|}{Charmonium $1^{-+}\ -\ 1S$}\\ \hline
 \multicolumn{2}{|c|}{Ref.~\& Method} & $N_f$ & $M$ (GeV) &
 \multicolumn{2}{|c|}{Ref.~\& Method}
  & $\Delta M$ (GeV) \\ \hline\hline
UKQCD 97\cite{ukqcd97}&SW&0&1.87(20)   &MILC 97\cite{milc97}     &W  &1.34(8)(20)\\
MILC 97\cite{milc97}  &W &0&1.97(9)(30)&MILC 99\cite{milc98}     &SW &1.22(15)   \\
MILC 99\cite{milc98}  &SW&0&2.11(10)   &CP-PACS 99\cite{cppacs99}&NR &1.323(13)  \\
LaSch 99\cite{lasch98}&W &2&1.9(2)     &JKM 99\cite{jkm}         &LBO&1.19       \\ \hline
\end{tabular}
\end{center}
\caption[tabB]{
  Recent results for the light quark and charmonium $1^{-+}$ hybrid meson
  masses.  Method abbreviations: W = Wilson fermion action; SW = improved
  clover fermion action; NR = nonrelativistic heavy quark action. $N_f$
  is the number of dynamical light quark flavors used.
  \label{table:light}}
\end{table}

\end{document}